# Automated Grain Boundary (GB) Segmentation and Microstructural Analysis in 347H Stainless Steel Using Deep Learning and Multimodal Microscopy

Shoieb Chowdhury[1], M.F.N. Taufique[1], Jing Wang[1], Marissa Masden[1], Madison Wenzlick[2,3]

Ram Devanathan[1], Alan L Schemer-Kohrn[1], Keerti S Kappagantula[1]

[1]Pacific Northwest National Laboratory, Richland, WA, 99354, USA

[2]National Energy Technology Laboratory, 1450 Queen Ave SW, Albany, OR, 97321, USA

[3]NETL Support Contractor, 1450 Queen Ave SW, Albany, OR, 97321, USA

**Abstract:** Austenitic 347H stainless steel offers superior mechanical properties and corrosion resistance required for extreme operating conditions such as high temperature. The change in microstructure due to composition and process variations is expected to impact material properties. Identifying microstructural features such as grain boundaries thus becomes an important task in the process-microstructure-properties loop. Applying convolutional neural network (CNN) based deep-learning models is a powerful technique to detect features from material micrographs in an automated manner. In contrast to microstructural classification, supervised CNN models for segmentation tasks require pixel-wise annotation labels. However, manual labeling of the images for the segmentation task poses a major bottleneck for generating training data and labels in a reliable and reproducible way within a reasonable timeframe. Microstructural characterization especially needs to be expedited for faster material discovery by changing alloy compositions. In this study, we attempt to overcome such limitations by utilizing multi-modal microscopy to generate labels directly instead of manual labeling. We combine scanning electron microscopy (SEM) images of 347H stainless steel as training data and electron backscatter diffraction (EBSD) micrographs as pixel-wise labels for grain boundary detection as a semantic segmentation task. The viability of our method is evaluated by considering a set of deep CNN architectures. We demonstrate that despite producing instrumentation drift during data collection between two modes of microscopy, this method performs comparably to similar segmentation tasks that used manual labeling. Additionally, we find that naïve pixel-wise segmentation results in small gaps and missing boundaries in the predicted grain boundary map. By incorporating topological information during model training, the connectivity of the grain boundary network and segmentation performance is improved. Finally, our approach is validated by accurate computation on downstream tasks of predicting the underlying grain morphology distributions which are the ultimate quantities of interest for microstructural characterization.

1. Introduction

Austenitic stainless steels [1-3] are pervasively used in extreme operating conditions including high temperature, high pressure components of fossil energy power plants. The efficiency of these power plants highly depends on the available materials. Computational modeling provides unique opportunities to devise new high temperature resistant materials [4, 5]. When designing heat resistant alloys for high temperature applications, one of the challenges is to tailor a microstructure that is stable in the power plant environment to increase the lifetime of the alloy. Various microstructural features contain valuable information and connections with their chemical and mechanical properties. Defects such as grain boundaries act as weak spots for corrosion and fatigue fracture [6-8], which are expected to be more influential at high temperature. For designing new materials suitable for extreme operating conditions, knowing the grain boundary structure is thus an important step in the process-microstructure-property loop.

Usually, scanning electron microscopes (SEM), scanning tunneling electron microscopes (STEM) and transmission electron microscopes (TEM) are widely used to study the microstructural and atomic level features of a given material. Until recently, most imaging techniques have relied on semiqualitative analysis, where human experts interpret two-dimensional (2D) images or individual one-dimensional (1D) spectra. However, data collected by electron and probe microscopes are often intrinsically quantitative and encoded across various modalities and dimensionalities. This characteristic requires subsequent extraction of features and correlations at various lengths and time scales. Additionally, data interpretation has been largely driven by human insights. This includes identifying features in images or spectra and qualitative interpretation, in some cases followed by quantitative analysis, ultimately connecting results to physical models and prior knowledge. This approach is inherently limited by human perception and bias. For example, the human eye is remarkably good at identifying well-localized objects but struggles to detect the emergence of correlated signatures in different parts of the image field or to detect small or gradual changes in periodicity. Furthermore, the eye is extremely sensitive to color scales and can regularly be deceived by the perception of contrast. Even more importantly, interpretation of the data in terms of relevant physics is highly dependent on prior knowledge [9, 10].

Advancements in computational tools, accessibility to experimental testing data and the growing field of integrated computational materials engineering (ICME) provide a useful basis for the application of data science and machine learning for optical and electron microscopy images. For instance, deep learning models were implemented to identify surface defects in steels [11]. Convolutional neural network (CNN) models have shown unprecedented success for microstructural characterization such as classification of several steel materials from microscopic images [12, 13]. Deep CNN models were applied to characterize different phases, such as martensite, ferrite and pearlite in low carbon steels [14]. DeCost et al. utilized ultrahigh carbon steel microstructures to calculate important feature vectors for segmentation tasks [15]. Deep convolutional neural networks have been successfully used to segment microstructural constituents of ultrahigh carbon and ferrite-martensite dual phase steels from microscopic images [16, 17]. To the best of our knowledge there is no significant literature on automated image segmentation for 347H stainless steel, which is an important candidate material for extreme environment applications. With that motivation under the Pacific Northwest National Laboratory (PNNL) and National Energy Technology's (NETL) effort on Extreme Environment Materials (XMAT) research, a database was developed comprising of many relevant data streams, including SEM and EBSD images of 347H stainless steel manufactured via casting and rolling at different processing conditions. In this work, we utilized the database to build several deep learning models for microstructural analysis by detecting grain boundaries and resulting grain morphology directly from SEM images. One existing drawback in using supervised deep learning models for microstructure analysis is that they require manual labeling for training. For segmentation tasks, generating training labels is expensive as hand annotations by human experts is needed. Additionally, it is often subjected to bias which negatively impacts the training data. We propose to combine SEM and EBSD images to directly generate labels for training deep learning models as a solution to that problem. This research critically investigates the advantages and limitations associated with such multi-modal microscopy approach by using several popular deep learning models for segmenting grain boundaries. We found that applying CNN and label generation by multi-modal microscopy showed great promise in automated microstructure segmentation and predicting grain size and shape from micrographs. Although, the choice of material is 347H stainless steels, the data generation, model training, and performance evaluation used in the current work can easily be applied and extended to other similar materials as well.

## 2. Methods
### 2.1. Micrograph collection

The specimens for this study are several 347H stainless steel alloys with composition and process variations that were provided by NETL. The specimens were cast with varying boron compositions. Once the cast samples were cooled down, they were rolled into thick sheets. Following this, specimens were obtained from each of the rolled sheets for microstructural characterizations. The specimens were mounted in epoxy and polished with silica. They were characterized in a JOEL 7600F SEM equipped with a high-speed Oxford Symmetry EBSD detector. EBSD data was analyzed using the Oxford AZtec Nanoanalysis Suite software. EBSD mapping of the specimens was done with an accelerating voltage of 20 kV, at a working distance of 24.5 mm, and a tilt of 70°. Only the steel matrix was mapped for these images to obtain the grain boundaries. Indexing was performed using a cubic crystal system, with 3.6599 Å lattice parameter. EBSD data post-processing was performed using Oxford AZtecCrystal software. The grain size was evaluated using the linear intercept method, and grain differentiation was performed at a misorientation of 2° and 10°. The images used were acquired from the forward scattering detector.

### 2.2. Image pre-processing and data augmentation

A total of 298 SEM images were collected for this study out of which 269 images had a size of 2048 ×1536 pixels (width vs height) and the rest (29) were of size 2048×1408 pixels. Each SEM image wass split into smaller non-overlapping square patches of 128 ×128 pixels to create the image dataset. Each image patch is then divided into either train/test (or validation) sets randomly with a 90-10% probability. This strategy ensured stratified sampling on the test set from specimens with different conditions. Five different deep learning model architectures, namely DefectSegNet, HED, UCF, U-Net and DeepCrack were first trained, and their performance was evaluated on this dataset of image size 128 by 128 pixels. A dataset with a larger 512 ×512 image size was created in a similar manner to train the best model architecture identified in the previous step and for downstream microstructural analysis. Figure 1-a shows an example SEM image with grain boundary overlayed acquired using EBSD. Figure 1-b shows the process of dividing the original SEM image into a grid of 512 × 512 image patches. All the SEM images were pre-processed before feeding into the deep learning models. The electron images were median filtered to remove any local noise [18]. To increase the size of training set and reduce possible overfitting, training data were augmented to four times in size by flipping the images (and labels) along the horizontal, vertical, and both horizontal and vertical axis. Additional data augmentation in the form of rotations in $90^0$ intervals were also performed but did not improve the model performance. No data augmentation was applied to test images. Table 1 shows the size of train and test sets for different image patches considered.

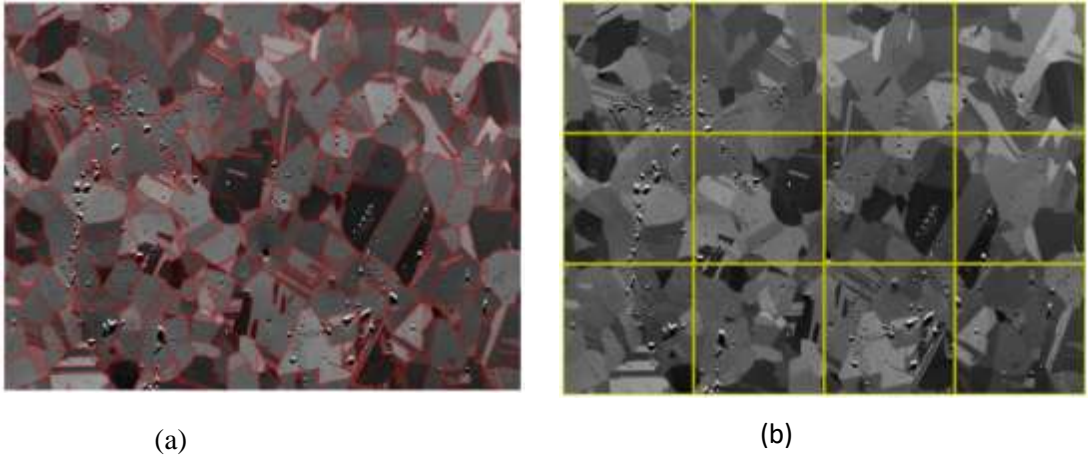

(a)           (b)

**Figure 1: a) A sample SEM image with the GB map overlayed in red b) Dividing a large SEM image into smaller size (512 x 512) patches to obtain stratified sampling in the train and test set.**

**Table 1 Size of training and test sets**

| Image Patch size | Training set size (after 4x augmentation) | Test set size |
|---|---|---|
| $128 \times 128$ | 1,98,288 | 5,566 |
| $512 \times 512$ | 12,464 | 344 |

## 2.3. CNN model architectures

We considered five different CNN model architectures designed for applications that are conceptually closely similar to the task of grain boundary segmentation. The most popular model considered was the U-Net architecture originally developed for biomedical applications [19]. It contains a contract path known as an encoder to learn context of images and a symmetric expanding path (decoder) for feature localization. The output features from each stage of the encoder are concatenated with the input to the decoder for better propagation of feature information. The model has demonstrated great success in applications like segmentation of neuronal structures and cell tracking with limited data. In this study, the code from reference [20] was used for implementing the U-Net architecture in Pytorch [21] and was trained on the available SEM images, and corresponding grain boundary maps of the 347H steel acquired using EBSD. The second model tested was the DefectSegNet [22], which features an encoder-decoder architecture with building blocks of DenseNet [23] where the output from one convolutional layer was fed as inputs to all the other downstream layers. The model was initially designed to predict microstructural defects in transmission electron microscopy images. This model combines the idea of U-Net and DenseNet architecture where not only each dense block but also the feature maps with the same spatial resolution across the encoder and decoder are connected. The model was re-implemented using Pytorch following the approach provided in [24]. The third and fourth models were developed specifically for edge detection. Grain boundaries are thin features in microscopic images akin to edges in natural images; therefore, we hypothesized these architectures would be suited for grain boundary detection. The third model explored was Holistic-nested edge detection (HED), which uses fully connected neural networks to learn hierarchical representation of edges [25]. The input image is processed at multiple scales and the results are then fused into a final output. Compared to conventional edge detection methods such as Canny edge detector [26], this method is typically more robust, suitable for learning hierarchical representation of edges, and faster. Similar to HED, the fourth model used in this study, Richer Convolutional Features (RCF) model fuses features at different stages; however, in contrast to HED, multiscale hierarchy is used to enhance edges in RCF [27]. HED and RCF models were implemented in Pytorch following [27, 28] and were trained using the EBSD + SEM images of 347H steels alone. Like edges, other type of thin features found in natural images are cracks in roads, pavements, walls etc. The final GB detection model we have tried is thus based on crack detection application called DeepCrack [29] and for this model both training from scratch and pre-trained weights in the form of transfer learning were adopted as the model trained parameters were readily available [30].

### 2.4. Model training

Deep learning CNN models were initially trained with a small image patch size of 128×128 using Pytorch framework with a batch size of 128 images. Training was carried out on PNNL's Institutional Computing Cluster using NVIDIA P100 GPUs in most cases when GPU memory permitted. For the DeepCrack model which has ~30.9 million parameters (which is considerably larger than the other models used in this study), two RTX 2080 Ti GPUs with 384GB memory were used for model training. Identifying grain boundaries in an SEM image is a highly class-imbalanced problem since grain boundaries class accounts for <10% of all pixels in an SEM image, where the rest is made of the grain class. Therefore, we used a balanced binary cross entropy loss (BCE) function with higher weight placed on the grain boundaries class. The weight ratio between two classes were kept consistent at 9.0 among different models for comparison. For the best performing model architecture (discussed in a subsequent section), the weight ratio between the classes was modified to observe its effect on model performance. ADAM optimizer [31] with a learning rate of $3e^{-4}$ was used across all the models. Early stopping of the training procedure was adopted to prevent overfitting. Other commonly used regularization methods such as Dropout [32] and L2 regularization [33] were also adopted with no noticeable trend of impact on the model performance.

After the five CNN models (DefectSegNe, HED, RCF, U-Net, and DeepCrack) were trained, their performances were evaluated on test set images (train and test images of size 128 by 128 pixels) using various metrics as discussed in the next section. The best CNN architecture (among the five models considered) for the pixel-wise grain boundary segmentation task was then selected and trained in a similar manner and its performance was tested on a dataset having image size of 512×512 pixels. The larger image size was selected to ensure that each image contained a sufficient number of grains for grain size and shape statistics calculations. Additionally, the pre-trained model was further trained to improve boundary connectivity. A topological loss function [34, 35] was added to the previously discussed BCE loss to further train the best model. The topological loss function is known to overcome the topological connectivity issues that arise from the naïve per-pixel cross-entropy loss. The dual loss function is shown in equation 1.

$$Loss_{dual} = Loss_{BCE} + \lambda\, Loss_{Topo} \qquad (1)$$

where the $Loss_{BCE}$ is the same BCE loss used before, and the $\lambda\, Loss_{Topo}$ is the topological loss with λ being a hyperparameter that controls the relative weight between the two losses. An ADAM optimizer with a learning rate of $1e^{-5}$ and λ = 0.04 was used to train the dual loss function. The trained models using both BCE and combined loss function steps were then used to report grain statistics calculations.

### 2.5. Performance metrics

To evaluate the performance of a CNN model, a few pixel-wise as well as overall metrics were calculated from the predicted binarized grain boundary map (pixel-wise classification of test set images into grain or grain boundary class). These include metrics that are often used in classification such as true positive rate (TPR), true negative rate (TNR), precision, and recall defined as following:

$$TPR = \frac{True\ Positive}{True\ Positive+False\ Negative} \quad (2)$$

$$TNR = \frac{True\ Negative}{True\ Negative+False\ Positive} \quad (3)$$

$$Precision = \frac{True\ Positive}{True\ Positive+False\ Positive} \quad (4)$$

$$Recall = \frac{True\ Positive}{True\ Positive+False\ Negative} \quad (5)$$

Positive and negative class in this context means a pixel contains grain boundary and grain interior, respectively. Thus, true positive or true negative is the count of actual grain boundary and grain pixels respectively that have been correctly predicted by the CNN model. False positive and false negative quantifies the number of grain boundary and grain pixels wrongly classified by the model respectively. For a global summary of segmentation performance, the dice index which measures the localization ability of the predicted map compared to the ground truth can be described as follows:

$$dice\ index = \frac{2*True\ Positive}{2*True\ Positive+False\ Negative+False\ Positive} \quad (6)$$

As a microstructure consists of several grains, pixels belonging to each grain in a micrograph can be considered as a cluster. A cluster evaluation metric known as adjusted rand index (ARI) was also reported [36] by identifying grains from both predicted and ground truth binary images (each pixel is either 0 or 1) from the test set.

### 2.6. Grain statistics calculation

Two major grain statistics of interest are grain size and morphology distribution. The individual grain area in both model prediction and ground truth labels were measured as the number of pixels belonging to each grain identified by a connected-component analysis algorithm available in scikit-image [37]. The size of a grain was then defined as the diameter of a circle that has an equivalent area. For grain morphology calculation, the pixels within a grain were fitted to a principal component analysis with n = 2 components [38]. The square root of ratio between explained variances by the first two principal components provides the aspect ratio of each grain where each grain is considered an ellipse in 2-D space. All grain statistics calculations were performed on the test dataset containing image patch size 512×512 pixels.

### 3. Results and Discussion

Performance of the five models on test images in terms of various metrics are shown in Table 2 along with the parameter size of each model. Recall that these models were trained on image patch size of 128×128 with BCE loss of a class weight ratio of 9 between the grain boundary and grain classes.

**Table 2: Performance comparison between different model architectures.**

| Model Architecture | Total parameters (millions) | TPR | TNR | Precision | Recall | Dice Score |
|---|---|---|---|---|---|---|
| DefectSegNet | 0.35 | 0.747 | 0.791 | 0.251 | 0.747 | 0.376 |
| HED | 14.72 | 0.734 | 0.792 | 0.249 | 0.734 | 0.372 |
| RCF | 14.803 | 0.732 | 0.802 | 0.258 | 0.732 | 0.382 |
| **U-Net** | **17.26** | **0.794** | **0.804** | **0.276** | **0.794** | **0.409** |
| DeepCrack | 30.903 | 0.761 | 0.806 | 0.27 | 0.761 | 0.398 |

The five different CNN architectures considered for this work, namely DefectSegNet, HED, RCF, U-Net, and DeepCrack were developed for varied applications. Among these, HED, RCF, and DeepCrack architecture were designed and optimized for natural images. DefectSegNet was specifically designed for microscopic images. As a general trend, larger models tended to perform better (except the DeepCrack model). Overall, the U-Net architecture with 17.27M parameters, performed the best on all the metrics showed in Table 2. U-Net was developed to work with limited data and designed for biomedical applications such as segmenting MRI, X-Ray as well as microscopy images [39, 40]. Its encoder-decoder style has been shown to generalize better in segmenting microscopic images of different materials [41] and performs according to its reputation in our task of grain boundary segmentation. It should be noted that U-Net, despite performing better, had fewer parameters than the much larger DeepCrack model. It is likely because the model architecture for DeepCrack was optimized for natural images. DefectSegNet with significantly lower number of parameters (as shown in Table 3), performs slightly better in terms of precision, recall, and dice-score than the HED architecture, pointing toward the need for architecture optimization for micrograph segmentation using CNN-based deep learning approaches. As U-Net performed the best amongst all the CNNs explored in this study, it was used for further improvement by adding topological connectivity and grain statistics calculation.

**Table 3: Effect of weight factor used for GB class used in model (U-Net) training on performance metrics. Inset shows the corresponding performances when trained on 512×512 size image patches.**

| Weight ratio (for positive class) | TPR | TNR | Precision | Recall | Dice Score |
|---|---|---|---|---|---|
| 4 | 0.615 (0.611) | **0.912** (0.906) | **0.395** (0.377) | 0.615 (0.611) | **0.481** (0.466) |
| 6 | 0.732 | 0.862 | 0.332 | 0.732 | 0.456 |
| 9 | 0.794 (0.784) | 0.804 (0.782) | 0.276 (0.251) | 0.794 (0.784) | 0.409 (0.38) |
| 11 | **0.825** | 0.773 | 0.255 | **0.825** | 0.389 |

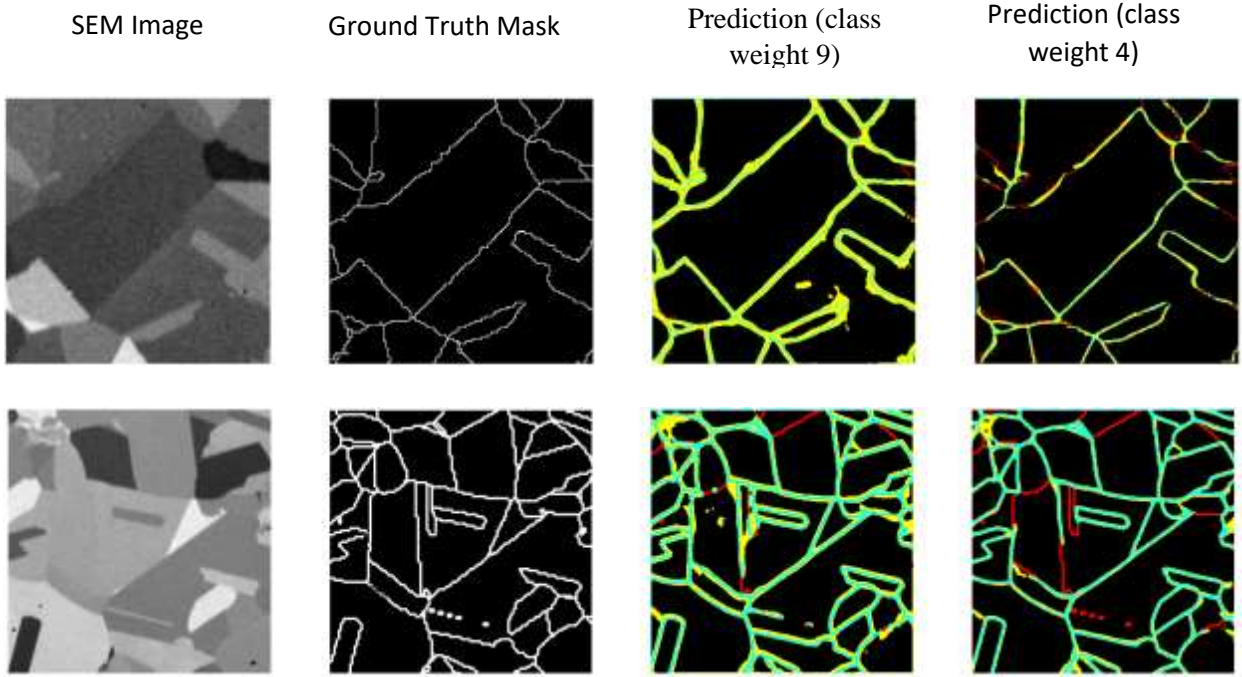

**Figure 2: Comparison of some images from test set between original SEM image, ground truth label, and the model prediction made by U-Net. The prediction image pixels are color coded as following: True negative = Black, True Positive = Cyan, False Positive = Yellow, False Negative = Red.**

As there is high class imbalance for the grain boundary pixels across all images, we have used a balanced cross-entropy loss by putting more weight on the grain boundary (positive) class. Experimentation with this weight ratio between positive grain boundary class and background grain class showed a trend in precision and recall performance as shown in Table 3 for U-Net model trained on image patch size 512×512. With increased positive class ratio, model recall also increased, but at the expense of precision. The highest precision found was 0.395 for a class weight ratio of 4 but this case also showed the lowest recall (0.615). On the other hand, a weight ratio 11 produced the highest recall (0.825) with the lowest precision (0.255) among all. This result indicates that a trade-off between precision and recall would be needed based on the desired goal. If the goal is to detect as many as the boundaries correctly (avoiding false negative), then a higher weight ratio would be needed. Conversely, if it is required to avoid false positives (higher precision), a lower weight ratio will increase the number of missed boundaries. Lower weight ratio between classes resulted in better Dice score (highest 0.481) and the performance was comparable (in some cases better) than a Dice score (0.433) reported for similar grain boundary segmentation task in unirradiated $LiAlO_2$ material [42]. For image size to 512×512, the trends still hold albeit with the performance dropping by a small margin as expected due to the increased size of the test images.

An accuracy plot (in terms of pixel-wise confusion matrix) for two sample images from the test set is shown in Figure 2, which once again highlights the trade-off between precision and recall as a function of class weight ratio variations. Large positive class weight (9) both qualitatively and quantitively missed fewer boundaries (less false negatives as shown in red). Identifying all the boundaries correctly is beneficial for the downstream task of grain statistics calculation as even a small, missed boundary section can lead to 'opening of the grain' and subsequently, inaccurate grain size, and shape calculation. Large positive weight, however, also overestimates the width of the boundaries (hence more false positives and the reduction in precision) and the curves in the grain boundary are smoother. This is partly due to the variable pixel width of grain boundaries present in the ground truth labels produced from EBSD, but the primary reason is the spatial instrumentation drift produced between the two modes of microscopy i.e SEM and EBSD, which will be discussed in the subsequent sections. In general, Figure 2 highlights that the trained CNN model using data generated from multimodal SEM and EBSD microscopy is successful in a challenging task of segmenting grain boundaries from SEM images. Even for an expert microscopist, manual labeling for grain statistics calculations become a highly tedious, time-consuming task, which can be accelerated with the use of CNNs developed in this study. By applying CNN segmentation model, the task of microstructure analysis directly from SEM images can be automated. As we have demonstrated for 347H stainless steel by segmenting micrographs with varying alloying composition, this method can significantly expedite the selection of alloys with desired microstructure. The CNN based semantic segmentation architectures tested in this study do not prioritize that the edges between grains should be closed or connected. Additionally, they do not explicitly differentiate between individual grains either as in the case of instance segmentation tasks [41]. This leads to small gaps in the predicted grain boundary network which produce large inaccuracies in downstream computation of grain statistics such as grain size and shape distribution. The conventional cross entropy loss function used here prioritizes pixelwise accuracy at the possible expense of the correct connectivity of the grain boundary which is a problem for microstructural characterization tasks. Although we have seen that a larger number of grain boundaries can be detected by increasing the class weight ratio, the effectiveness approach can work only so far before being overtaken by the false positives. Thus, there is a need to segment with correct topology. Following the work of Hu et. al. [34, 35], a loss function that prioritizes both pixel-wise accuracy and topological similarity of the segmented binary image with the ground truth, namely TopoLoss, was thus used for improving the connectivity of predicted microstructure. This method utilizes persistent homology and Morse theoretic measures defined on the grain boundary probability surface to capture topological information [42, 43] rather than trying to maximize the accuracy of each individual pixel on its own. We analyze the impacts of using TopoLoss to emphasize the appropriate connectivity of predicted microstructures. Figure 3 shows the segmentation accuracy predicted by the U-Net model on a couple of test images. By using the standalone BCE loss function, we can see the model predicted that the microstructures contain small gaps highlighted by orange markers, that would not impact the overall pixel-wise accuracy for segmentation task on natural images. But, for the task of automated computation of grain characteristic distributions and analyzing the impact of process and alloying elements on the resulting microstructure, that would be deemed insufficient. By considering topology of the ground truth labels, immediate improvements in connectivity and closing of grain contours are observed. Global performance metrics such as Adjusted rand index (ARI) [36] that reflects boundary connectedness also improves from 0.72 to 0.78 pointing to improved segmentation of grains and their boundaries as presented in Table 4.

**Table 4** Cluster similarity score in terms of adjusted rand index (ARI) shows improvement in grain boundary connectivity by the inclusion of topological loss function.

| Loss Function | ARI |
|---|---|
| BCE | 0.72 |
| BCE + TopoLoss | 0.78 |

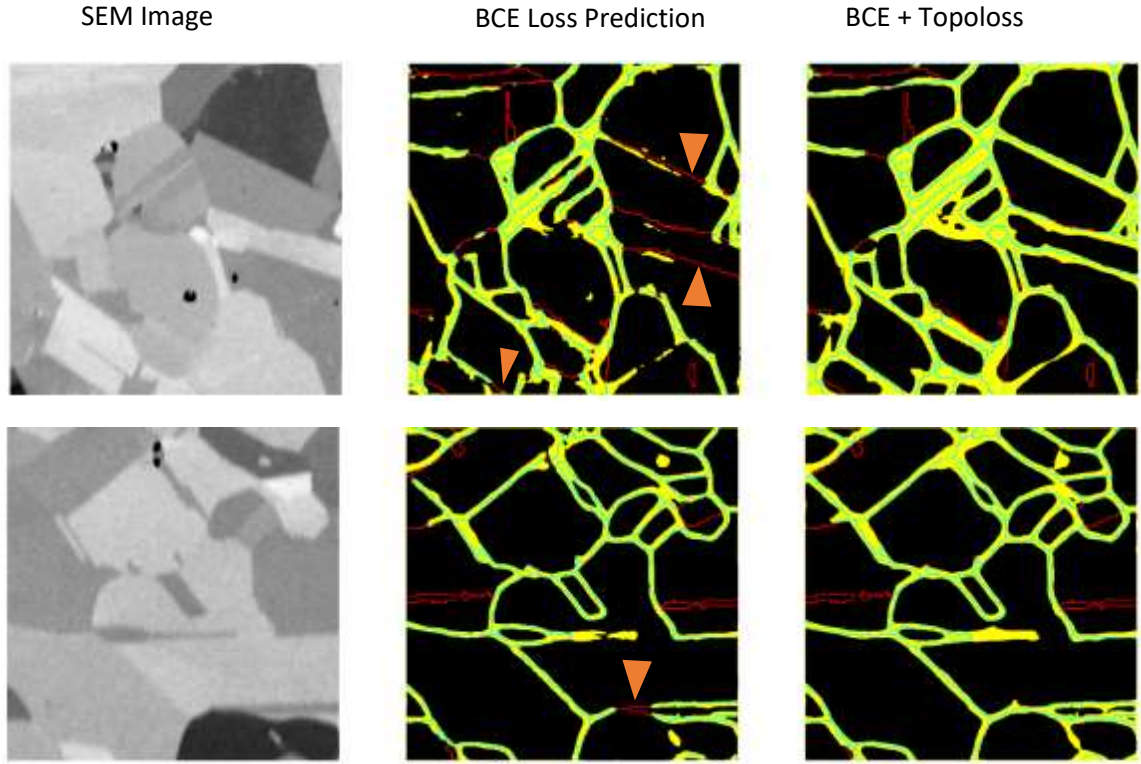

**Figure 3.** Pixel-wise prediction accuracy plot showing improvement in GB connectivity when a topological loss function is added to the pixel-wise BCE loss. Some small gaps in GB prediction maps (orange markers) are successfully being closed by considering topological connectivity during model training.

As the performance of any machine learning model is heavily reliant on the quality of data available for training and CNN models being no different, we discuss some aspects of using multimodal microscopy as a method for generating both the images and labels. During acquisition of high magnification SEM images as in the present case, there have been reports of drifts or spatial movement of pixels in the image [44, 45]. The factors that influence such pixel drifts include imaging time, charge build up, and sample preparation among others. Depending on SEM magnifications used and image acquisition time, there has been reported drift of up to 42 pixels [45] in some instances. As we used both SEM and EBSD microscopy modes that usually involve different image capture times [42], the amount of pixel drift between two images is inequal. As a result, we end up having a relative drift between the SEM image and its corresponding grain boundary map generated from the EBSD image. An example of such relative drift is presented in Figure 4, where with human-eye both the images and the binary segmentation mask look the same. Only when they are overlaid, the relative drift becomes apparent. For pixel wise segmentation models based on CNN, this poses a problem where the model faces a dilemma between its actual captured features from the image and what it sees from labeled segmentation mask where there are arbitrary pixel shifts. The outcome is that the model ends up overestimating the width of boundaries in its predictions in an effort to maximize the pixel-wise accuracy. Some of the CNN models such as HED and RCF have been reported to predict wider boundaries [25, 46]. However, U-Net and DeepCrack with their symmetric encoder and decoder structure are known to generate thin boundaries [29]. We find that all of the considered models, predict wider grain boundaries than ground truth. Additionally, we notice that if the U-Net model is overfitted to the training data, it is able to generate thin boundaries by memorizing the ground truth labels along with their pixel drifts rather than actually extracting relevant features from the images. With this evidence, we arrive at the conclusion that the existence of pixel-drift is the main reason why the model predicts wider grain boundaries compared to the ground truth. It should be noted that it is possible to reduce the extent of such drift either during the microstructural imaging [47] or during post-processing [48]. However, an important goal of this work is to evaluate multimodal microscopy as an automated approach for training CNN models. Besides, as we have variable amount of drift due to sample preparation history, as is often the case in materials microscopy, the task of drift correction becomes extremely challenging and would be an independent work on its own.

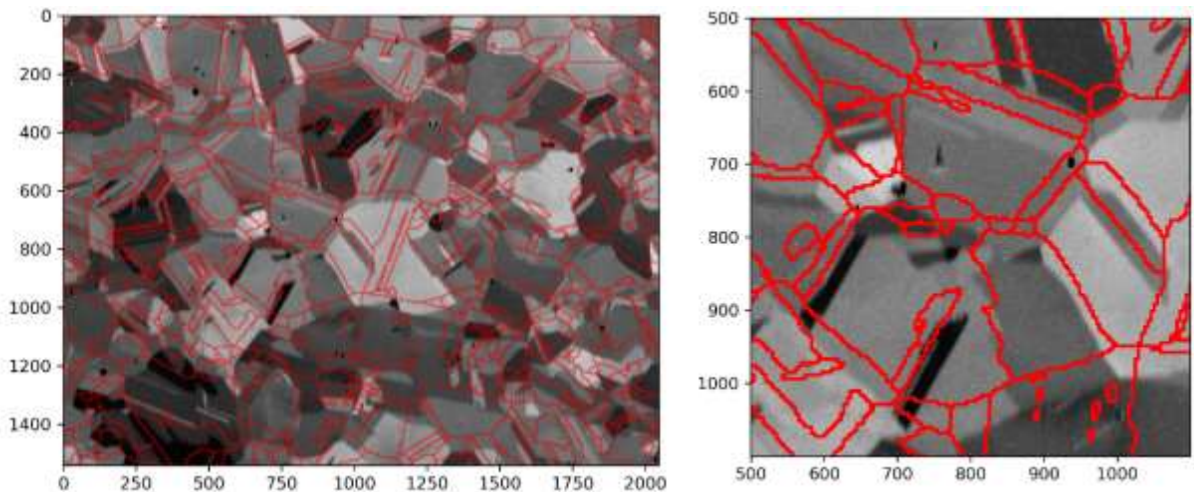

**Figure 4: An example SEM image with GB label overlaid highlighting the pixel drift found in some images. Inset shows a magnified view in a small region. Axes labels in the images show no. of pixels to act as a visual guide for understanding the extent of relative pixel drifts.**

The existence of the variable image drifts in the captured images warrants further quantification of the model-predicted grain boundaries beyond just reporting of conventional metrices such as precision, recall, and dice score. Figure 5 shows that the grain boundaries in the ground truth labels are only a few pixels wide. The variability in pixel width used (highlighted by the bi-modal distribution with two distinct peaks, (Figure 5-a)) for creating the grain boundary labels contribute to some extent in wider boundary predictions made by the CNNs.  In the case of model predictions, the boundaries are consistently wider (Figure 5-b) than the labels primarily due to the underlying relative drifts between images as discussed previously. On average the model predicted boundaries are 9 pixels wider (Figure 5-c) with a maximum value as large as 21 pixels wider than the ground truth. Note that the grain boundary width difference in the model predictions is within the range of relative spatial drifts such as shown as an example in Figure 4. These calculations show that despite the model precision being low when predicting the grain boundary class, the false positives mainly arise from the underlying pixel drifts present between the image and labels. Wider boundary predictions result in reduced precision (as shown in Table 3) on test set, however the model can still reliably segment grain boundary and grains for microstructural analysis.

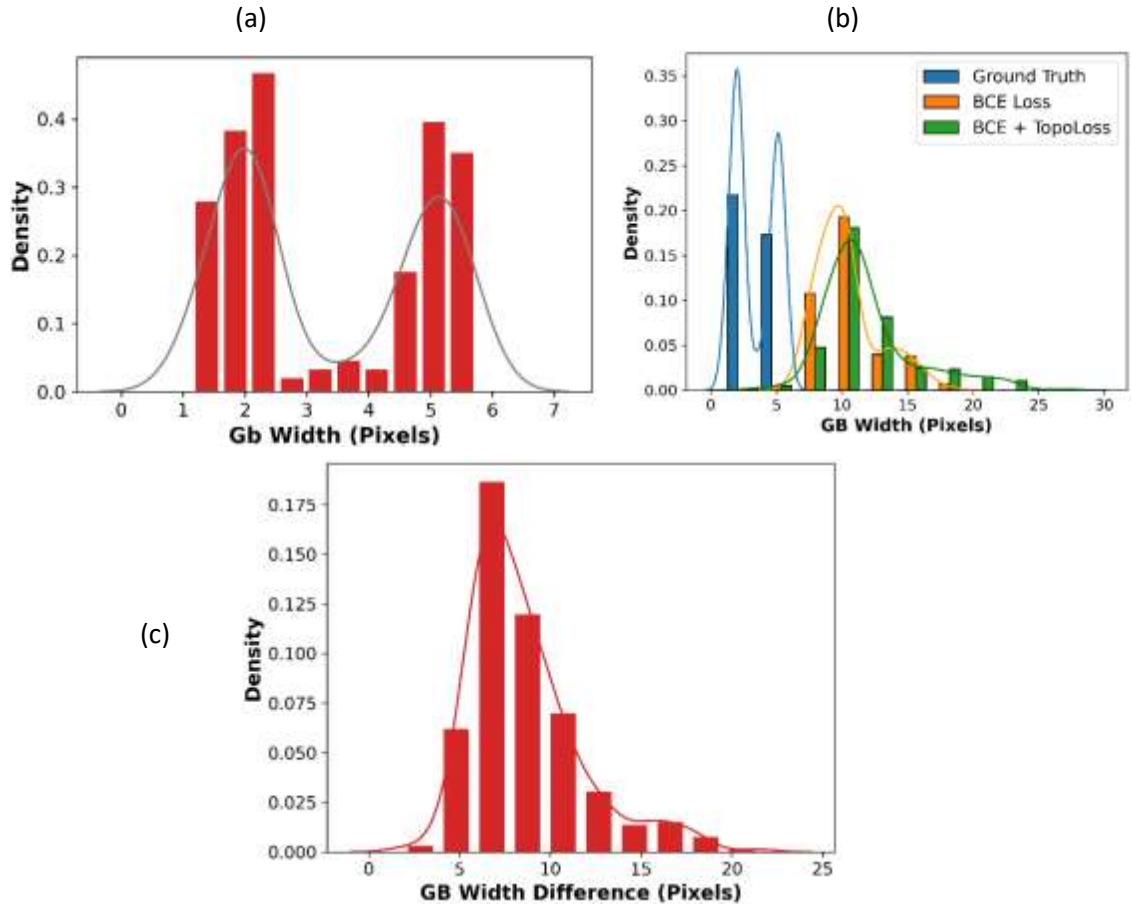

**Figure 5: Calculation of grain boundary width (a) Distribution in Ground truth labels created from EBSD images (b)Width of boundaries predicted by model shows a distribution that is right shifted (c) Difference between the predicted and true labels shows the extent of overestimation of boundary width by the CNN model.**

### 3.1. Microstructural analysis: Grain morphology

Macroscopic behavior of polycrystal material quantified as mechanical, electrical, wear, thermal, corrosion properties among others largely depends on the microstructural features of the material such as distribution of grain size, and grain morphology [49-51]. Additionally, microstructural features provide insight into the history of the manufacturing, thermal, and mechanical processes the material has undergone [52, 53]. Automated segmentation of grain boundaries allows subsequent characterization of material microstructure in terms of grain size and morphology both during and after such processes have occurred.

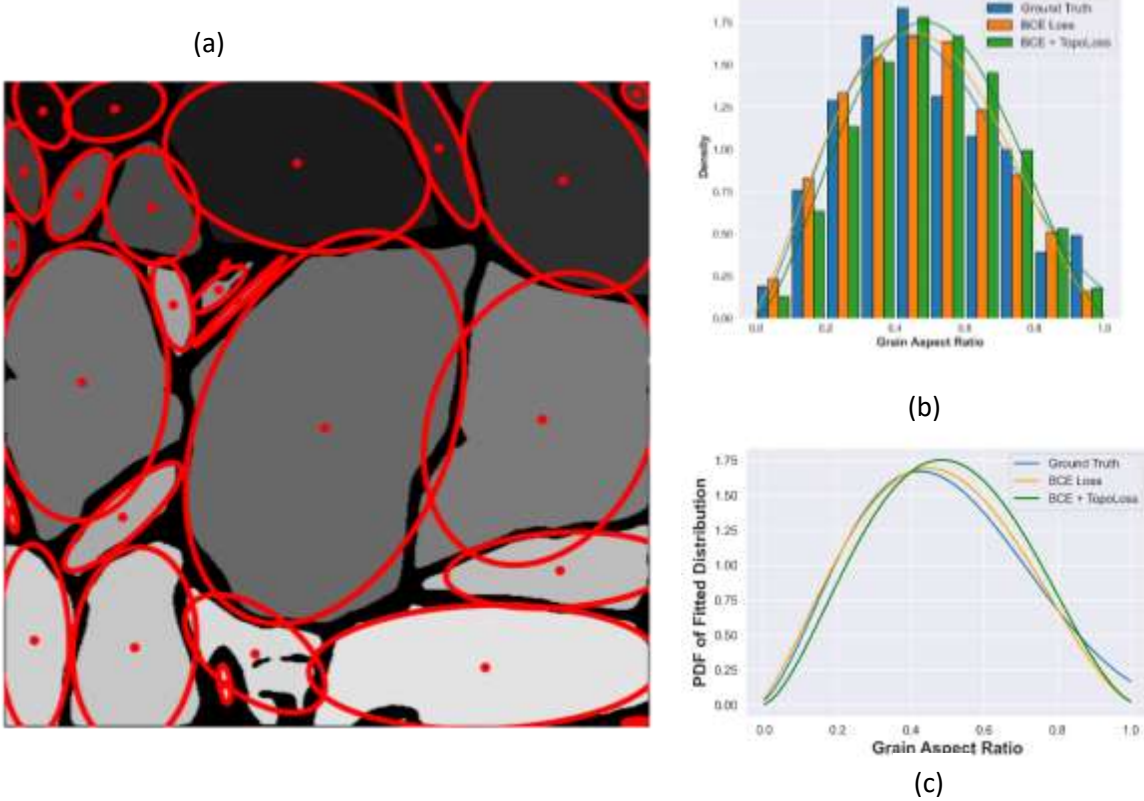

**Figure 6: (a) Grains as ellipses and their aspect ratio calculation as a shape parameter (b) Grain shape distributions extracted from EBSD ground truth and model predictions across test set images (c) Comparison between true and predicted shape distributions after Beta PDF fitting.**

In the subsequent section we show that grain size and shape calculations on the model segmented microstructures still are reliable despite having precision that stems from the pixel drift during the image acquisition step. To derive the microstructural features of interest, first, we calculated the grain shape (aspect ratio) across the microstructural images in the test dataset. Figure 6 presents an analysis of the grain shape in terms of their aspect ratios across all the images in the test set. As shown in Figure 6-a, only a few portions of the grains are equiaxed; most of the grains are elliptical with some being thin grains. The histogram distribution of all the grains across images obtained from EBSD ground truth also shows the same trend (Figure 6-b) and highlights that the existence of some extremely thin and tall grains. Variation in grain shapes is potentially due to the chemical variations in the 347H steels being analyzed in this study. To evaluate the ability of the U-Net for microstructural characterization, grain shapes were also calculated from the model predicted grain boundary binary images. We find that the distribution of grain shapes predicted by the model (both trained with BCE loss and Topological loss) closely follows the true distribution. The grains were identified using a connected component analysis in sci-kit image which is sensitive to small gaps in the predicted grain boundary map. This led to larger grains being predicted than the original case for some situations and the predicted distribution of grain shapes shifted slightly to the right as a result. However, the model still performed well to capture the underlying distribution of grain shapes and can be used to extract microstructural information directly from a SEM image. To compare the grain shape distributions, we fit a beta probability distribution to both the true and predicted grain shapes shown in Figure 6-c. As grain aspect ratio is always between 0 and 1, it is a natural choice adopted in other studies as well [54]. A similarity metric between two probability distributions can be calculated using the Jensen–Shannon distance [18]. The Jensen–Shannon distance between the true and predicted distribution is small (0.043 for BCE loss and 0.062 for Topological loss used), indicating the two distributions are quite similar. Beyond the regular accuracy metrics, this analysis indicates that the model was successful in predicting the underlying 347H stainless steel grain shape distributions.

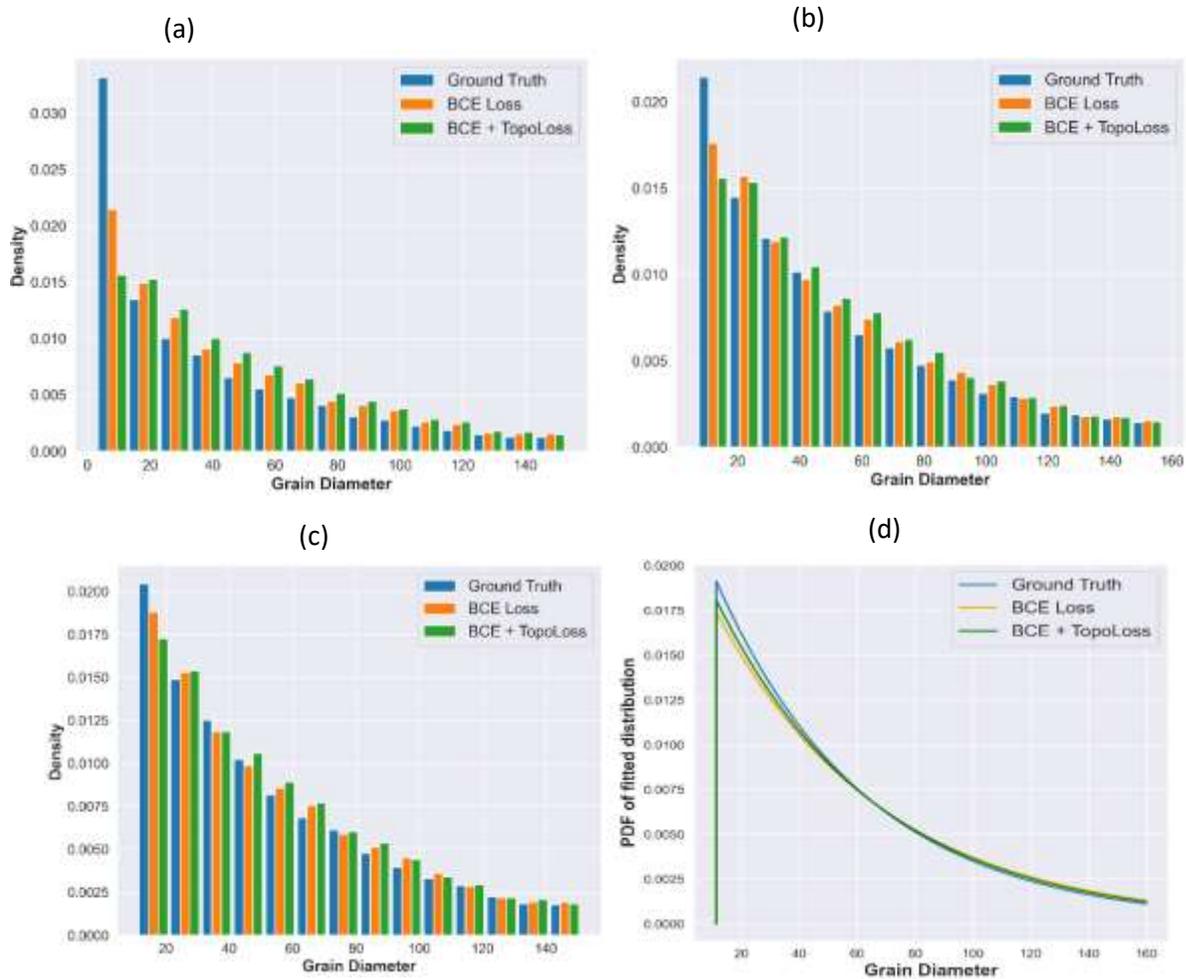

Figure 7. Grain size (circular equivalent diameter) distributions with minimum grain size of three, seven, and eleven pixels respectively (a-c). (d) Comparison between true and predicted grain size (minimum of 11-pixel size) distribution after exponential PDF fitting.

Grain size is another microstructural feature of interest in this study as it is an indicator for properties such as hardness, corrosion behavior, and yield strength. Both SEM and EBSD images show several twin grain boundaries. These twin boundaries are at 60° misorientation angle and are usually very thin. The true distributions of grains in Figure 7-a shows that a significant portion of grains are of small size and likely consist of complete and incomplete twin boundaries. Compared to the actual distributions obtained from EBSD segmented grain boundaries, the model overpredicts large size grains. This is due to the existence of small gaps in predicted boundary maps that do not fully enclose a grain resulting in two small grains merging into a larger grain spuriously. Due to the existence of instrument pixel drift between image and labels as discussed previously, we calculated that the model predicts on average 9-pixel wider grain boundary. Consequently, the model struggles to resolve the small sized twin-like boundaries and underpredicts the small grains as shown in Figure 7 a-b when detecting minimum grain size of 3- and 7-pixel diameters respectively. If the analysis is limited to minimum grain diameter of 11 pixels, we find the model predicted distribution closely matches the true grain size distribution (Figure 7-c). For comparing the prediction capability, an exponential probability distribution function is fitted to grain sizes calculated from EBSD and model predicted results in Figure 7-d. The Jensen–Shannon distance between the true and predicted distribution is calculated to be 0.0221 for BCE loss and 0.0136 for Topological loss used respectively, which is very low and indicates that the predictions are very similar to the ground truth data. Despite the limitation of resolving thin twin boundaries arising from pixel drifts during data collection, the CNN model trained on multimodal microscopy data still is quite successful in predicting the grain size distribution of 347H stainless steels for automated microstructural analysis.

4. **Conclusions**

In this paper, we presented an automated microstructural segmentation method for SEM images of 347H stainless steel using deep learning-based CNN models and computer vision. To facilitate the calculation of microstructural statistics (grain size and morphology), CNN models were trained to perform pixel-wise segmentation into grain boundary and grain interior classes. We demonstrated a novel multi-modal microscopy approach to circumvent the need of expensive manual pixel labelling by utilizing grain boundary mapping obtained through the corresponding EBSD image of each SEM micrograph. We evaluated several different CNN architectures with varying sizes of model parameters that were originally developed in different domains such as biomedical image segmentation, edge and crack detection in natural images, and defect segmentation of material micrographs. Generally, model architectures that were optimized for microscopy-like images performed better. The best-performing model architecture, U-net, trained on in-house SEM images of 347H steel with different compositions showed comparable performance on similar task that used manual labelling. We found that combining multi-modal microscopy data for generating both the images and labels has its challenge in the form of instrument induced pixel drift of spatial location of images as the two microscopy modes require different timescales. As a result, there is a mismatch between the feature extracted by the model from SEM images and the ground truth GB map in some cases. The deep learning model compensated for this by predicting wider grain boundaries and showed a trend in precision vs. recall trade-off depending on the weight applied to the grain boundary class. As it is a problem with high class imbalance (with <10% of the pixels in an image belonging to the grain boundary class), a high model recall compromised precision in the form of predicted grain boundaries being wider by 9 pixels on average. The width overestimation of the predicted grain boundaries, however, is within the maximum relative pixel drift present during the EBSD data capture, providing validation of the developed model's efficacy. As even a small gap in predicted grain boundary map can lead to inaccurate calculations of grain size and shape distributions, we devised a dual topological loss function along with the cross-entropy loss during model training which showed improvement in maintaining topological connectivity of GB network both qualitatively and quantitatively. Finally, our approach is shown to successfully calculate grain statistics from the predicted grain boundary map and lead to accurate measurement by comparing the original and predicted analytical distributions for grain size and shape.

**Acknowledgments**

This work was supported by the U.S. Department of Energy, Office of Fossil Energy, eXtreme Environment MATerials (XMAT) consortium. This research used resources of the Pacific Northwest National Laboratory (PNNL), which is supported by the U.S. Department of Energy. PNNL is operated by the Battelle Memorial Institute for the United States Department of Energy under contract DE-AC06-76LO1830. M.M. was supported in part by an appointment with the National Science Foundation (NSF) Mathematical Sciences Graduate Internship (MSGI) Program sponsored by the NSF Division of Mathematical Sciences. This program is administered by the Oak Ridge Institute for Science and Education (ORISE) through an interagency agreement between the U.S. Department of Energy (DOE) and NSF. ORISE is managed for DOE by ORAU.

**Author contributions**

S.C and M.F.N.T. conceived the study. J.W. and M.W. managed the dataset. S.C performed machine learning training and drafted the manuscript. M. M. implemented the topological loss function. A.L.S.K performed EBSD analysis. K.S.K., R.D., M.M., M.F.N.T., M.W. edited the manuscript.

**Competing interests**

The authors declare no competing interests.